# Quantitative Single-particle Profiling of Extracellular Vesicles via Fluorescent Nanoparticle Tracking Analysis

Yiting Liu[a], Anthony James El-helou[a], Bill Söderström[b], Juanfang Ruan[c] and Ying Zhu[*a,d,e,f]

[a] School of Biomedical Engineering, University of Technology Sydney, Australia

[b] Australian Institute for Microbiology and Infection, University of Technology Sydney, Australia

[c] Electron Microscope Unit, UNSW Sydney, Australia

[d] School of Clinical Medicine, Faculty of Medicine & Health, UNSW Sydney, Australia

[e] Institute for Biomedical Materials and Devices, University of Technology Sydney, Australia

[f] Australian Centre for NanoMedicine, UNSW Sydney, Australia

## Abstract

Extracellular vesicles (EVs) have drawn rapidly increasing attention as the next-generation diagnostic biomarkers and therapeutic agents. However, the heterogeneous nature of EVs necessitates advanced methods for profiling EVs at the single-particle level. While nanoparticle tracking analysis (NTA) is a widely used technique for quantifying particle size and concentration, conventional scattering-based systems are non-specific. In this study, we present an optimised protocol for quantitative profiling of EVs at the single-particle level by fluorescent NTA (F-NTA). The protocol integrates fluorescent immunolabeling of EVs with size-exclusion chromatography (SEC) to efficiently remove unbound labels, enabling the precise quantification of EV concentration, size distribution, and surface immunophenotype. We first validated this approach using biotinylated liposomes and EVs from cultured human cell lines, confirming effective removal of unbound labels and assessing labelling efficiency. We then demonstrated that F-NTA can distinguish EV subpopulations with distinct surface marker expression, exemplified by the differentiation of EpCAM-positive EVs derived from HT29 and HEK293 cells. Finally, we applied dual labelling to human plasma isolates to simultaneously profile EVs and non-vesicular extracellular particles, providing a quantitative quality assessment of EV purity at the single-particle level. The robustness of this method was further supported by comparative analysis with total internal reflection fluorescence microscopy. This validated workflow enables robust, quantitative profiling of EV subpopulations, providing a critical tool for diverse EV applications, including biomarker discovery, therapeutic monitoring, and quality control for engineered vesicles.

## Introduction

Extracellular vesicles (EVs) are lipid bilayer-encapsulated micro- to nanoscale particles secreted by all cell types. EVs have drawn rapidly increasing attention as the next-generation diagnostic biomarkers and therapeutic agents, due to their critical roles in intercellular communication.[1-3] Their molecular cargo and surface markers, comprising proteins, lipids, and nucleic acids, can reflect the physiological or pathological state of the originating cells, making them especially valuable for non-invasive diagnostics. However, EVs are inherently heterogeneous, owing to diverse biogenesis pathways and cellular origins.[4] EVs derived from the same parent cell type can encompass multiple, distinct subpopulations. This complexity presents a major challenge for conventional bulk analytical methods such as Western blotting and enzyme-linked immunosorbent assay (ELISA), which lack the resolution to distinguish individual vesicle subtypes. Consequently, there is a critical need for single-EV analysis technologies capable of probing vesicles at the individual level, enabling more precise molecular characterisation and offering deeper insights into EV biology, function, and clinical potential.

The growing need for single-particle analysis of extracellular vesicles (EVs) has driven the development of a wide array of detection techniques and technologies.[5, 6] Among the numerous single-EV analysis techniques, nanoparticle tracking analysis (NTA) remains the most widely adopted, owing to its ability to simultaneously quantify size distribution and particle concentration at the single-particle level, while also being widely accessible and easy to use through commercially available platforms. NTA is an optical technique that tracks the Brownian motion of individual particles in suspension to estimate their hydrodynamic diameter via the Stokes–Einstein equation, while also quantifying particle concentration based on detected scattering events. However, conventional NTA systems rely primarily on total light scattering, rendering the measurements non-specific. This limitation hampers molecular discrimination and raises ambiguity as to whether the detected particles are indeed EVs or co-isolated contaminants.

To address this need for specificity, fluorescent NTA (F-NTA) has emerged as a powerful advancement. F-NTA enhances the conventional scattering-based NTA configuration through the addition of optical bandpass filters, which enable the selective detection of fluorescently labelled particles. F-NTA has enabled fluorescence-based phenotyping of EVs at the single-particle level, offering a means to evaluate labelling strategies and molecular markers with greater specificity. For instance, Fortunato et al. utilised F-NTA to evaluate several fluorescent dyes, including lipid and antibody dyes to label EVs, as well as different washing methods to remove the free dyes.[7] Mladenović et al. recently compared the performance of F-NTA in quantitative fluorescence analysis with another single EV analysis technique, nano-flow cytometry, and determined that the limit of detection of F-NTA is ~21 Alexa Fluor molecules.[8] Dlugolecka et al. provided one of the first examples of using F-NTA to analyse EVs from patient samples using the bronchopulmonary lavage fluid, and suggested the challenges of using F-NTA to measure complex fluids.

In this study, we provided an optimised protocol for using F-NTA to profile EVs carrying different surface phenotypes by measuring EVs labelled with fluorescent antibodies. We found that the effective removal of unbound fluorescent antibodies using size-exclusion chromatography (SEC) is crucial to a successful F-NTA measurement. While many studies have used the ultracentrifugation method to pellet EVs together with unbound fluorescent labels, this has been reported in a previous study to potentially even contribute to the problem of micelle creation.[9] To avoid these issues, we opted for SEC, which enables gentle and efficient separation of particles from the unbound label. F-NTA for measuring fluorescently labelled nanoparticles was first validated with biotinylated liposomes labelled with Alexa Fluor™ 488 (AF488) streptavidin, as well as EVs labelled with AF488 anti-CD63. Both F-NTA measurements showed the expected labelling efficiency, calculated by the proportion of fluorescent particles among total particles. F-NTA successfully differentiated between EVs carrying different levels of surface markers, demonstrated by measuring AF488 anti-EpCAM labelled EVs derived from HT29 and HEK293, which have known differences in EpCAM expression from our previous work.[10] Finally, we demonstrated that the F-NTA can profile different types of extracellular particles from human plasma isolates by specifically labelling EVs with Alexa Fluor 647 (AF647) anti-CD63 and lipoproteins with AF488 anti-ApoB, which shows the capability of F-NTA to assess the quality of EV isolation at the single-particle level. This quantitative quality assessment was also compared with another single EV technique using total internal reflection fluorescence (TIRF) microscopy imaging.

**Experimental**

**Materials**

SEC qEV1 70 nm column and Automatic Fraction Collector (AFC) V2 were purchased from Izon Science (USA). Non-PEGylated biotinylated liposomes (CDEIMS-1523), DiO-labelled liposomes (CDL6001F-DO) and DiD-labelled liposomes (CDL6001F-DD) were purchased from CD Bioparticles (USA). Lyophilised exosomes from HT29 (HBM-HT29-100/2) and HEK293 (HBM-HEK293-100/2) cell culture supernatant were purchased from HansaBiomed Life Sciences (Estonia). Very low-density lipoprotein (437647-5MG) was purchased from Merck Life Science (Germany). BD Vacutainer K2E EDTA tube (367525) was purchased from Becton Dickinson (UK). AF488 streptavidin conjugate (S11223) and AF488 anti-CD63 (MA5-18149) were purchased from Thermo Fisher Scientific (USA). AF488 anti-ApoB (sc-393636 AF488) was purchased from Santa Cruz (USA). Both AF488 anti-EpCAM (ab237395) and AF647 anti-CD63 (ab309976) were purchased from Abcam (UK). Phosphate-buffered saline (1× PBS) tablets (P4417) and Poly-L-lysine (PLL) solution (P8920) were purchased from Sigma-Aldrich (USA). Bottle-top vacuum 0.22 μm filter system (CLS430517) was purchased from Corning (USA). Particle Metrix Zetaview PMX-420 QUATT and polystyrene (PS) 100 nm standard calibration beads were from Particle Metrix (Germany). High-precision coverslips No. 1.5, 25 × 75 mm (GVD7247) and sticky-Slide VI 0.4 (80608) for TIRF imaging were purchased from Knittel Glasbearbeitungs (Germany) and ibidi (Germany), respectively. The Nikon Super-resolution microscope N-STORM was from Nikon (Japan).

**Fluorescence labelling of EVs, lipoproteins and liposomes**

EVs, lipoproteins and liposomes were labelled with fluorescent antibodies or proteins according to the conditions in Table 1. Specifically, the sample and fluorescence label under the listed concentrations and volumes were mixed and incubated at room temperature in the dark with continuous low-speed (250 rpm) shaking for 2 hours. All necessary dilutions were performed with 0.22 μm filtered particle-free 1× PBS.

*Table 1: Summary of fluorescence labelling conditions for liposomes, lipoproteins and EVs.*

| Sample | Sample Vol. (µL) | Particle concentration (/mL) * | Fluorescence label | Label Vol. (µL) | Label Vol. (mg/ml) |
|---|---|---|---|---|---|
| Biotin-liposome | 100 | $1.6 \times 10^9$ | AF488 Streptavidin | 5 | 0.5 |
| Lipoprotein | 100 | $2.2 \times 10^{10}$ | AF488 anti-ApoB | 10 | 0.2 |
| HT29 | 100 | $6.6 \times 10^{10}$ | AF488 anti-CD63 | 10 | 0.26 |
| HT29 | 100 | $6.6 \times 10^{10}$ | AF647 anti-CD63 | 10 | 0.25 |
| HT29 | 100 | $6.6 \times 10^{10}$ | AF488 anti-EpCAM | 10 | 0.25 |
| HEK293 | 100 | $1.8 \times 10^{10}$ | AF488 anti-EpCAM | 10 | 0.25 |
| Human plasma isolate | 200 | $6.2 \times 10^{10}$ | AF647 anti-CD63 | 20 | 0.25 |
| | | | AF488 anti-ApoB | 20 | 0.2 |

*The particle concentration was measured by scattering NTA measurement.

**Size-exclusion chromatography (SEC) for unbound label removal and plasma EV isolation**

The qEV1 70 column was loaded in AFC and used based on the manufacturer's instructions. In general, the column was first washed with 27 mL particle-free 1×PBS before sample loading. The sample was topped up to 1 mL particle-free 1×PBS for column loading if the original volume was below 1 mL. A total volume of 2.8 mL was collected after the buffer volume. After every SEC round, the column was washed with 13.5 mL 0.5M NaOH and 27 mL of 1×PBS.

The same qEV1 70 nm column was used for both purposes of unbound label removal and plasma EV isolation, with slightly different buffer volumes based on the manufacturer's user manual and recommendations. For unbound label removal, the buffer volume was set to 4 mL, so that the sample was collected from 4.0-6.8 mL to maximise the EV recovery. For plasma EV isolation, the default buffer volume of 4.7 mL was used, and the samples were collected from 4.7-7.5 mL to maximise the EV purity.

**Nanoparticle tracking analysis (NTA)**

Particle concentration and size distribution were measured by Zetaview PMX-420 QUATT in scatter and fluorescence modes with software version 8.05.16_SP3. The instrument was auto-aligned with the 100 nm PS standard beads before sample measurement. Before each sample injection, the flow cell was flushed with fresh Mili-Q water and then primed with fresh PBS to avoid turbulent drift. Samples were diluted with particle-free 1×PBS to reach the optimal particle concentration within the manufacturer-recommended measurement range (approximately $5 \times 10^6$ to $1 \times 10^8$ particles/mL).

The detailed parameters for scattering and fluorescent NTA measurements are listed in Table 2. The default laser wavelength for scattering measurement was 488 nm. For fluorescence measurement, laser/filter wavelength combinations 488/500 nm and 640/660 nm were used for AF488 and AF647 antibodies, respectively. All fluorescence measurements were conducted under the "Low Bleach" mode to minimise photobleaching during data acquisition. In case of multiple measurements of a single sample, the "Multiple Acquisitions" and "Dose Sub Volume" modes were used to enhance statistics of the replicate measurements. For each measurement, 11 different positions across different focal planes of the flow cells were measured, generating 11 replicates for the particle concentration and size distribution data, including median and mode sizes. For size distribution, a bin size of 10 nm was used for the histogram display.

Table 2. The detailed parameters for scattering and fluorescent NTA measurements.

| Type | Pre-acquisition parameter | | | | Post-acquisition parameter | | | |
|---|---|---|---|---|---|---|---|---|
| | Camera sensitivity | Camera shutter | Number of frames | Frame rate | Minimum Brightness | Min Area | Max Area | Tracelength |
| Scattering (S-NTA) | 80 | 100 | High | 30 | 30 | 10 | 1000 | 15 |
| Fluorescent (F-NTA) | 95 | 100 | Low | 30 | 30 | 10 | 1000 | 7 |

For each fluorescent measurement, commercial fluorescent liposomes that matched the specific laser/filter wavelengths were used as the fluorescence standards to obtain the concentration correction factor. DiO liposomes were used for 488/500 nm, and DiD liposomes were used for 640/660 nm. Briefly, the fluorescence standards were measured first in both scattering and fluorescence modes. A number of Particles vs. Sensitivity (NvS) graph was generated to plot the relationship between sensitivity and the number of particles detected in the field of view. The concentration correction factor ($K_f$) was determined by the following equation:

$$K_f = \frac{n_{det}(S, sens = 80)}{n_{det}(F, sens = 95)}$$

Where $n_{det}(S, sens = 80)$ is the number of particles detected in scatter mode at a sensitivity of 80, and $n_{det}(F, sens = 95)$ is the number of particles detected in fluorescent mode at a sensitivity of 95. All fluorescence concentrations were then multiplied by the concentration correction factor to obtain the corrected concentration values.

**Human blood collection and plasma extraction**

Blood samples were collected from the Australian Red Cross (Life Blood Sydney, NSW, Australia) using 10 mL EDTA tubes (UTS Human Research Ethics Committees Approval ETH21-5782). Plasma was extracted by a two-step centrifugation according to a previous methodological guideline to reduce platelet-derived EVs.[11] Briefly, the whole blood was first centrifuged at 2500 × g for 15 minutes at room temperature. The supernatant was gently transferred to a new 15 mL tube and centrifuged again at 2500 × g for 15 minutes. The collected plasma was aliquoted into 1 mL and stored at -80°C before use.

**Total Internal Reflection Fluorescence (TIRF) microscopy imaging and data processing**

The fluorescently labelled human plasma isolation sample was immobilised onto a coverslip for TIRF microscopy imaging. A coverslip was cleaned with $O_2$ plasma for 1 minute, incubated with 1 mg/mL poly-L-lysine (PLL), rinsed with water, and blow-dried. The PLL-coated coverslip was assembled with a 6-channel sticky slide cartridge (sticky-Slide VI 0.4, ibidi). 30 μL sample was introduced to the cartridge channel and incubated in the dark for 1 hour, which allows the electrostatic adsorption of the negatively charged particles (EVs and lipoproteins) to the positively charged coverslip. Then, the channel was rinsed with 120 μL of 1×PBS. After washing, 60 μL of 1×PBS was added to each reservoir, and the cartridge was transferred to the microscope for imaging.

Fluorescence images were acquired on a Nikon TiE2 N-STORM microscopy equipped with a sCMOS camera and a CFI HP Plan Apochromat VC 100× oil immersion objective, operated in TIRF mode. Low laser power of 5% and an exposure time of 400 ms were used for both 488 nm and 647 nm laser filter settings. The Perfect Focus System (PFS) was activated to maintain the samples in focus throughout the image acquisition time. The region of interest (ROI) was set to 1024 × 1024 pixels (0.065 μm/pixel). To ensure representative sampling while minimising photobleaching, images were captured at eight distinct locations spaced from one side of the channel to the other.

The fluorescence images were subsequently analysed using Fiji ImageJ. The open-source plugin, EVAnalyzer, was used for automatic quantification of EV numbers in each image.[12] The EVAnalyzer processes all images in the chosen input folder with the same settings and saves the results in a default, automatically created output folder. For a single experiment, all samples and controls were analysed with consistent settings. The specific parameter settings are as follows: Series to import: Series_1 (488nm channel) and Series_2 (647 nm channel); Function: EV Counts, Type: EV_GPF and EV_CY5; Threshold algorithm: Manual; Manual threshold: 100; Min Circularity: 0. The image analysis results consist of a report file in .xlsx format, output images, and a .json file, which includes all settings and can be used to reload the configuration.

**Data analysis**

All data were analysed and visualised using GraphPad Prism (Version 10.4.2). Bar charts were presented as mean (solid bar) ± standard deviation (vertical error bar). The violin plots were presented with medium smoothing. The median value of each group is indicated by a dashed line within the plot. Statistical comparisons between two groups were performed using an unpaired two-tailed t-test. P values were considered statistically significant as follows: $\leqslant 0.05$ (*), $\leqslant 0.01$ (**), $\leqslant 0.001$(***), $\leqslant 0.0001$ (****).

For the initial validation experiments (Figures 2 and 3), the SEC eluate (4.0–6.8 mL) was collected as four separate 0.7 mL fractions. Each fraction was measured once, generating data from 11 positions within the NTA flow cell. As no consistent trend was observed between the individual fractions, the data from all four fractions were pooled for analysis, resulting in a total of n=44 replicate measurements. For subsequent experiments (Figures 4 and 5), the entire 4.0–6.8 mL eluate was pooled first, and this single sample was measured in triplicate. With each measurement comprising 11 positions, this resulted in a total of n=33 replicate measurements.

**Results and Discussion**

Figure 1 presents the schematic for the whole workflow, including EV labelling, unbound label removal, F-NTA measurement and EV isolation from plasma. For fluorescent labelling, EVs or other investigated particles, such as lipoproteins and liposomes, were mixed with the fluorescent labels in a microtube at room temperature for two hours on a microtube shaker. The mixture was run through the SEC column using the AFC. The flow of smaller particles along the column is slower than that of larger particles, as they tend to be trapped by the porous resin, resulting in size-based particle separation. Labelled EVs or other particles were collected during the earlier stages of elution, leaving the unbound antibody or protein labels in the later stages. An elution volume 4.0-6.8 mL were collected as suggested by the manufacturer's recommendation to maximise the EV recovery (see Experimental for details).

After removing unbound labels, the purified EVs were measured by NTA. We use the Zetaview PMX-420 from Particle Metrix for NTA measurement, which can measure the total particles through scattered light as well as fluorescent particles through fluorescence emission. A 100% fluorescent particle usually appears with a higher particle count in the scattering NTA (S-NTA) measurement than in the F-NTA measurement. This is because weak fluorescent particles may not be detectable, as the emitted fluorescence light must pass through a filter before reaching the camera. To compensate for this loss, we used a sensitivity of 95 in fluorescence mode instead of 80 in scattering mode and corrected the fluorescent particle number using a concentration correction factor (see Experimental for details). The correction factor ($K_f$) was obtained by measuring fluorescent liposomes, which have a similar lipid bilayer structure to EVs and thus serve as a more accurate reference material compared to solid nanoparticles, such as fluorescent polystyrene beads. The fluorescent liposomes were formed by incorporating lipophilic dyes (DiO, DiD etc.) into the lipid bilayers during production, and thus can be considered 100% fluorescent particles.

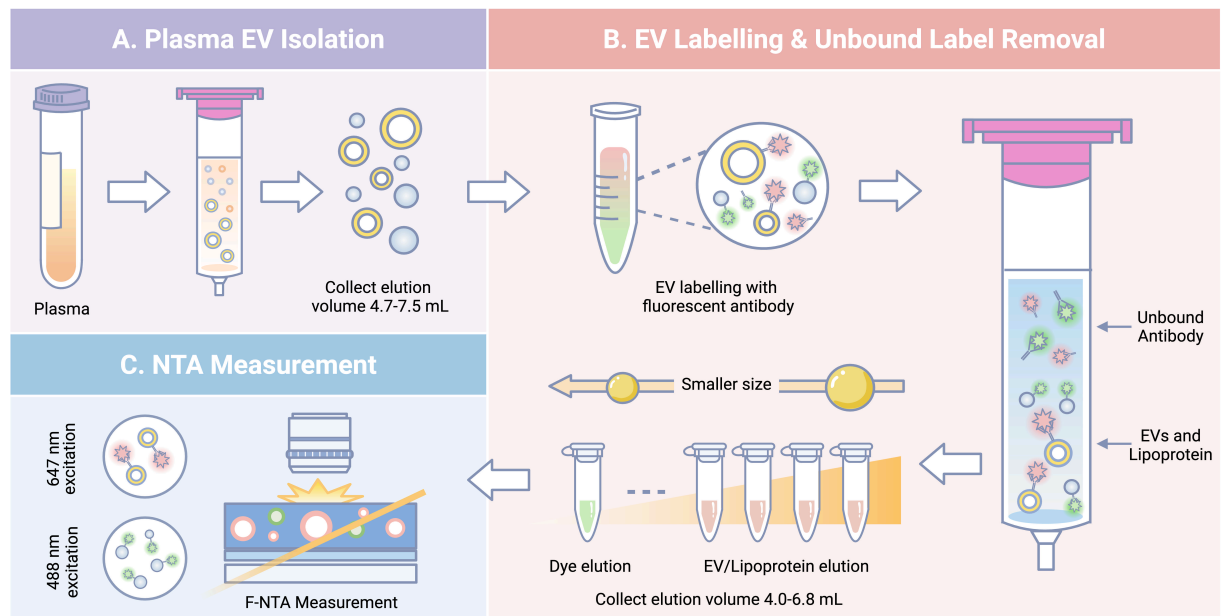

Figure 1. Schematic diagram of the whole F-NTA workflow. A) Plasma EVs are isolated using the SEC column. The elute volume of 4.7–7.5 mL is collected to maximise the EV purity. B) EVs and other investigated particles, like lipoproteins and liposomes, are labelled in solution with fluorescent antibodies or proteins. Different fluorescent dyes, including AF488 (green)

and AF647 (pink), are used for multiplexed detection. To remove excess unbound labels, the labelled samples were further processed through the same SEC column but collected at a different elution volume of 4.0–6.8 mL to maximise the EV recovery. C) F-NTA is then used to measure fluorescently labelled particles.

**SEC effectively removes unbound fluorescence labels**

A key challenge for F-NTA measurement is the interference from unbound fluorescence labels, which cause false-positive signals and high background noise, making weakly fluorescent particles undetectable. This challenge increases for EVs due to their small density, which makes it difficult to separate them from the fluorescence labels. Figure 2 shows that SEC effectively removes unbound fluorescence labels by comparing the fluorescence antibody solutions with the same concentration for labelling before and after SEC label removal. Figure 2A shows the field of view images from the F-NTA of an example solution, AF488 anti-ApoB, before and after SEC label removal. The result showed that the fluorescent intensity was dramatically reduced after the label removal process. This background reduction enabled the reliable measurement of fluorescently labelled EVs and other investigated particles, which are otherwise undetectable due to high background fluorescence from the fluorescent labels.

Subsequently, we quantified the number of fluorescent particles after labelling and SEC label removal and investigated the influence of any possible unremoved fluorescent labels. We used EVs derived from the human colon adenocarcinoma cell line HT29 (HT29 EVs) for demonstration. Figure 2B shows the results from HT29 EVs labelled with AF488 anti-CD63, compared with the same volume of solution that contains only the anti-CD63 but no EVs. The unremoved labels accounted for only 4.22% of the total fluorescent particles, which were considered negligible for our subsequent investigations. We also investigated very low-density lipoprotein (VLDL) from human plasma, which is a predominant type of particle from human plasma that co-isolates with EVs.[13] The VLDL was labelled with AF488 anti-ApoB and was investigated with the same comparison as HT29 EVs. Figure 2C shows comparable results to Figure 2B, in which the unremoved labels account for only 8.66% of the total fluorescent particles, demonstrating the versatility of the SEC process in removing different fluorescence antibody labels.

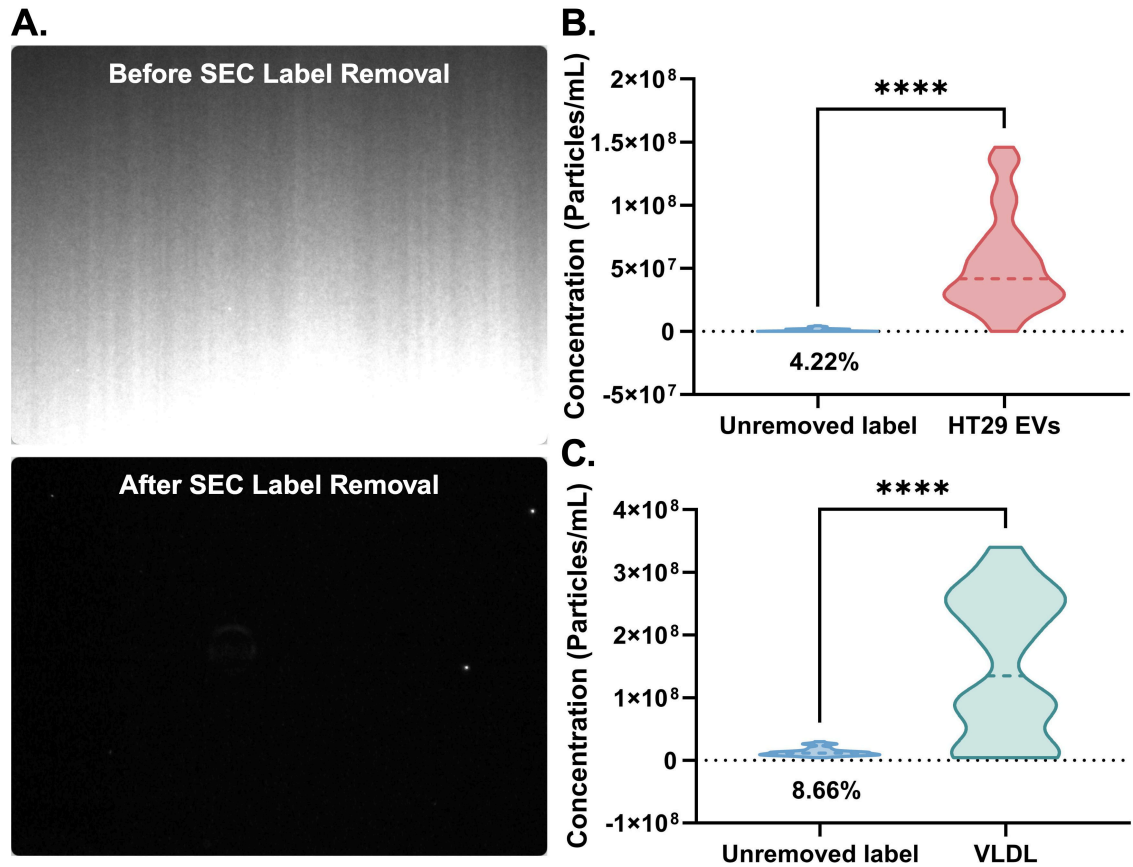

Figure 2. Performance of SEC label removal. A) Screenshots of the instrument-recorded videos before (top) and after (bottom) SEC label removal using AF488 anti-ApoB for demonstration. High background noise is observed in the original fluorescence antibody solution, which drops dramatically after the SEC label removal. B-C) Comparison of the particle concentrations of labelled HT29 EV (B) and VLDL (C) and their corresponding free labels after SEC label removal. The proportion of unremoved labels out of the total fluorescent particles is 4.22% for HT29 EVs and 8.66% for VLDL, which are considered negligible (p < 0.0001). The data are from n = 44, which includes a sum of four fractions (0.7 mL each) that were collected from the elute volume 4.0-6.8 mL and measured separately. Each sample was measured at 11 positions (see Experimental section for details). The median value of each group is represented as a dashed line.

**Validate particle labelling efficiency with fluorescent proteins and antibodies**

Next, we validated the fluorescent labelling efficiency using liposomes and EVs by F-NTA measurement after SEC label removal. The labelling efficiency was calculated as the ratio of the corrected fluorescent particle concentration ($C_{F\text{-}NTA}$) to the total scatter particle concentration ($C_{S\text{-}NTA}$), using the following equation:

$$Labelling\ efficiency\ (\%) = \frac{C_{F-NTA} \times K_f}{C_{S-NTA}} \times 100$$

Firstly, we investigated biotinylated liposomes labelled with AF488 streptavidin, due to the highly efficient and stable biotin-streptavidin binding. Figure 3A shows that the labelling efficiency is as high as 63.2%. Size comparison reveals no significant difference between the scatter and fluorescent modes, though the size distribution for F-NTA appears broader than that for S-NTA (Figure 3B). The apparent broadening of the size distribution in F-NTA compared to S-NTA may stem from heterogeneous labelling efficiency among EV subpopulations, which leads to tracking variations. Furthermore, a shorter minimum trace length, defined as the minimum number of consecutive frames in which a particle must be tracked to be included in the analysis, was used in F-NTA to reduce photobleaching, resulting in increased statistical uncertainty in particle size estimation. Labelling HT29-derived EVs with Alexa Fluor 488-conjugated anti-CD63 antibodies resulted in 8.09% of the total particles exhibiting detectable fluorescence, reflecting the effective labelling efficiency under the given experimental conditions. This proportion is considered acceptable given that only CD63 was targeted, and not all EVs within the heterogeneous population are expected to express this single marker.[14] Furthermore, this percentage is comparable to that reported previously in a similar work by Mladenović et al., which was approximately 2%.[8] Notably, cryo-electron microscopy of the commercially sourced HT29 EVs revealed the presence of non-EV contaminant particles alongside vesicular structures (Figure S1). This indicates that the fluorescent particle percentage measured by F-NTA is an underestimation of the labelling efficiency between anti-CD63 and CD63+ EVs, as the total particle number by S-NTA also contains non-EVs. Size comparison revealed a modest increase in the average particle diameter from 114.9 nm to 133.1 nm, consistent with observations by Fortunato et al., and attributable to the added hydrodynamic size contributed by antibody labelling.[7]

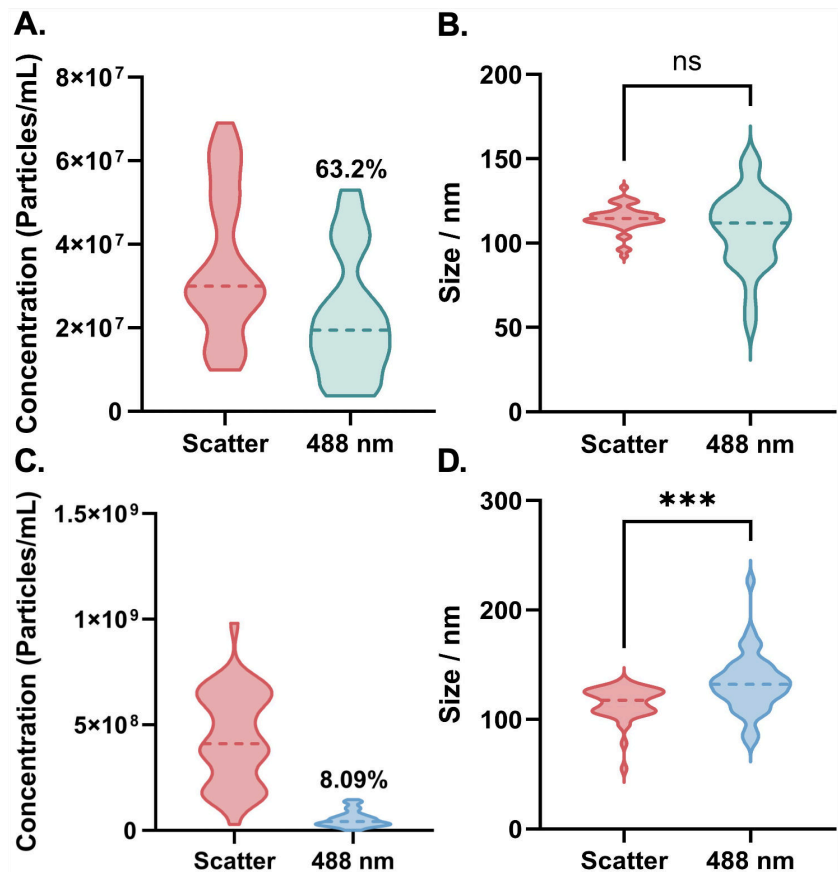

Figure 3. Particle labelling efficiency. A) Concentration and B) Size analysis of AF488 streptavidin-labelled liposomes measured in scatter (red) and fluorescent (green) modes. The proportion of fluorescent particles in total particles is 63.2%. A size comparison reveals no significant difference (p=0.0859) between the scatter and fluorescent modes. C) Concentration and D) Size analysis of AF488 anti-CD63 labelled HT29 EVs measured in scatter (red) and fluorescent (blue) modes. The proportion of fluorescent particles in total particles is 8.09%. Size comparison shows a slight increase in the average size from 114.9 nm to 133.1 nm (p=0.0002). The data are from n = 44, which includes a sum of four fractions (0.7 mL each) collected from the elute volume of 4.0-6.8 mL and measured separately. Each sample was measured at 11 positions (see Experimental section for details). The median value of each group is represented as a dashed line.

### F-NTA can differentiate between EVs carrying different levels of surface markers

We then investigated the potential of F-NTA to discriminate between EV subpopulations based on differences in surface marker expression levels.. Our previous study demonstrated that HT29 EVs contain a significantly higher proportion of EpCAM-positive (EpCAM$^+$) EVs compared to the human embryonic kidney cell line HEK293 (HEK293 EVs), providing well-defined comparison samples for validating F-NTA's capability for biomarker profiling.[10] HT29 and HEK293 EVs were labelled with AF488 anti-EpCAM and measured by NTA after SEC label removal. Figure 4A presents the particle concentrations from S-NTA and F-NTA for both EVs, and Figure 4B shows the corresponding fluorescent particle proportion comparison. The results clearly show that HT29 EVs carry a significantly higher proportion of EpCAM$^+$ EVs (61.63%) compared to HEK293 EVs (19.43%). This result also indicates that in a single EV analysis technique, such as F-NTA, it is not necessary to consider sample loading or establish calibration curves in bulk methods like Western blotting or ELISA, as the ratio between fluorescent particles and total particles can be used to quantify the biomarker difference. Size distributions show that the normal distribution is retained for both EVs after fluorescence labelling. While no significant size distribution difference was observed for HT29 EVs before and after labelling (Figure 4C), the size distribution of HEK293 EVs showed a shift towards the smaller size (Figure 4D). This shift may be attributed to the low fluorescent particles from the HEK293 EVs, which makes it hard to obtain a statistically robust size distribution.

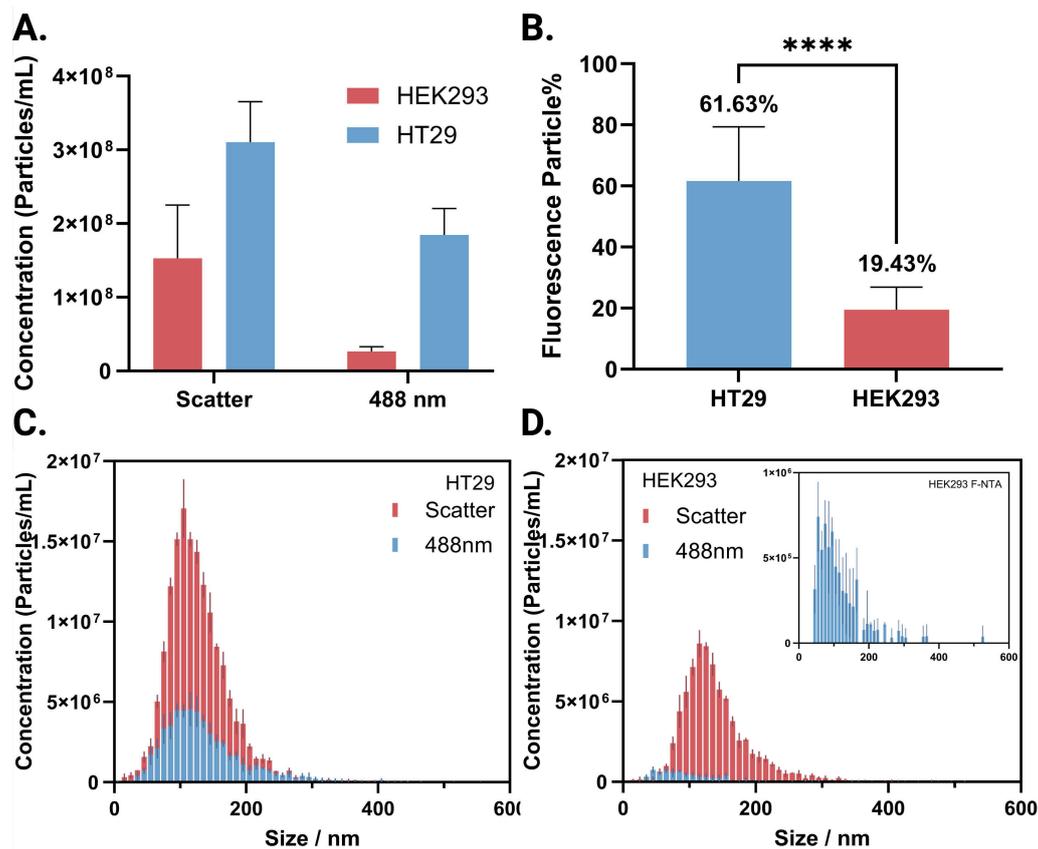

Figure 4. Comparison of EpCAM[+] EVs between HT29 and HEK293 EVs. A) Total particle concentrations from S-NTA (scatter) and fluorescent particle concentration from F-NTA (488 nm). B) Comparison of EpCAM[+] EVs, as shown by the proportion of fluorescent particles. HT29 EVs (61.63%) showed a significantly higher proportion (p < 0.0001) compared to HEK293 EVs (19.43%), consistent with the different EpCAM expression of the EVs from the two cell lines. The data are from n = 33, which includes three measurements of a sample at 11 positions from the total collected volume 4.0-6.8 mL (see Experimental section for details). (C-D) Corresponding size distributions from (C) HT29 EVs and (D) HEK293 EVs in scatter (red) and fluorescent (blue) modes. The size distribution histograms are presented as mean (solid bar) ± standard deviation (vertical error bar).

**F-NTA can be used to assess the purity of EVs isolated from complex samples**

The EV purity is an essential parameter for EV isolation. Obtaining highly pure EVs is challenging, particularly in complex samples like human plasma. In our previous study, we established a bulk method to assess EV quality using immunomagnetic beads EV capturing, followed by flow cytometry analysis.[13] Therefore, in this study, we investigated whether F-NTA can be used for quality assessment of plasma-derived EVs at the single-particle level.

For plasma EV isolation, the same SEC column used for label removal was employed, but with an adjusted elution volume of 4.7–7.5 mL to maximise EV purity (see Experimental section for details). Cryo-electron microscopy images confirmed the presence of EVs in the isolates (Figure S2). The isolates were labelled with two fluorescent antibodies with different fluorescent dyes, AF647 anti-CD63 for EV labelling, and AF488 anti-ApoB for lipoprotein labelling. Lipoproteins were used to assess the impurity in plasma EV isolation because they are the primary contaminant particles in human plasma, which are several orders of magnitude more abundant than the amount of EVs.[15] Anti-ApoB recognises non-vesicular extracellular particles (NVEPs), including VLDL and chylomicrons, which have similar densities and sizes to EVs and are hard to separate by SEC isolation methods. F-NTA measurement shows the presence of both EVs and NVEPs. Figure 5A shows the size distribution of the EVs and NVEPs, with the concentrations of NVEPs significantly higher than those of EVs across the interrogated sizes. The EV: NVEP ratio is 7.44% from the F-NTA measurement (Figure 5C). We compared the results obtained by F-NTA with another single EV technique using TIRF microscopy imaging. The same sample, measured by F-NTA, was introduced onto a PLL-coated microscopy coverslip through the electrostatic absorption of EVs and NVEPs, and then imaged by a fluorescence microscope using TIRF illumination. Figure 5B shows the presence of anti-CD63 labelled EVs (pink dots) and anti-ApoB labelled NVEPs (green dots) in the same field of view. A quantitative analysis of the image shows an EV: NVEP ratio of 27.92%. This difference in quantification can be attributed to the distinct detection principles of the two methods. F-NTA is a dynamic technique that analyses particles freely diffusing in solution. For a particle to be counted, it must produce a fluorescent signal bright enough to exceed the detection threshold within the short exposure time of a single video frame. This makes F-NTA a powerful tool for characterising particle size and concentration in solution, but sets a stringent requirement for signal intensity. In contrast, TIRF microscopy is a static method that analyses particles immobilised on a surface. This allows for the use of longer exposure times, enabling the accumulation of photons from even dimly labelled particles. While this approach offers high sensitivity for enumerating surface-captured fluorophores, it does not provide the in-solution hydrodynamic size and concentration data that F-NTA does. These fundamental differences in particle state (dynamic vs. static) and signal acquisition logically result in the observed quantitative discrepancies.

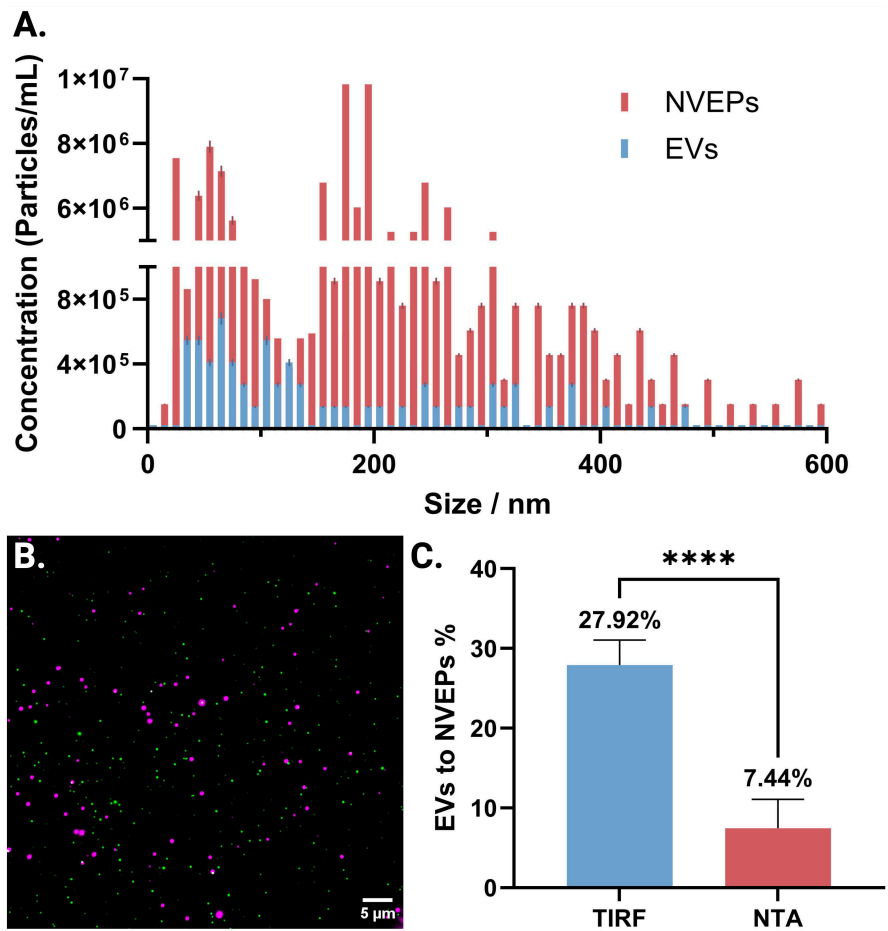

Figure 5. A) Concentration distribution of EVs and NVEPs measured by F-NTA across particle sizes ranging from 5 to 600 nm. The histograms are presented as mean (solid bar) ± standard deviation (vertical error bar). B) Representative TIRF image of labelled EVs (magenta) and NVEPs (green). Scale bar: 5 μm. C) Quantitative comparison of EV-to-NVEPs detected by TIRF and F-NTA. TIRF analysis revealed a significantly higher (p < 0.0001) EV-to-lipoprotein ratio (27.92%) compared to F-NTA (7.44%). The data from TIRF are from n = 8, obtained from eight different imaging locations across the microfluidic channel. NTA data are from n = 33, which includes three measurements of a sample at 11 positions from the total collected volume 4.0-6.8 mL.

## Conclusions

In this study, we established an optimised protocol for quantitative single-particle profiling that integrates fluorescent antibody labelling, SEC label removal and F-NTA. This robust workflow enables reliable quantification of surface marker expression and the accurate assessment of EV purity in a single, integrated approach. Looking forward, the capabilities of F-NTA can be further expanded. Future advancements, such as improving camera sensitivity, using next-generation ultra-bright fluorophores, and implementing advanced analytical software, will continue to improve the detection limit for dimly labelled and rare EV populations. Furthermore, this workflow provides a foundation for multiplexed co-localisation analysis using next-generation NTA instruments. Ultimately, this work provides the research community with a critical and accessible tool for developing EV-based diagnostics, monitoring therapeutic efficacy, and performing quality control on engineered vesicles for drug delivery.

## Author contributions

YL: conceptualisation, data curation, formal analysis, investigation, methodology, validation, visualisation, writing – original draft; AE: formal analysis, writing – review & editing; BS: resources, supervision, writing – review & editing; JR: investigation, writing – review & editing; YZ: conceptualisation, formal analysis, funding acquisition, methodology, project administration, writing – original draft.

## Conflicts of interest

The authors declare no conflicts of interest in this work.

## Data availability

The datasets supporting this article have been uploaded as part of the ESI.† Further information relevant to this study is available from the authors upon request. We have submitted all relevant data of our experiments to the EV-TRACK knowledge base (EV-TRACK ID: EV250078).

## Acknowledgements

We acknowledge the following funding support: the Australian Research Council Discovery Early Career Researcher Awards (DE240100321) for Y. Z. and the Australian Research Council Future Fellowship (FT230100062) for B. S.; PanKind, The Australian Pancreatic Cancer Foundation (24.R10.EDG.YZ.UTSY) for Y. Z.. We thank Dr Sven Kreutel from Particle Matrix for his technical support. We also thank Dr Yunyun Hu and Professor Xiaomei Yan from Xiamen University for their discussions regarding antibody dye removal and EV labelling.

# Supplementary Information

## Quantitative Single-particle Profiling of Extracellular Vesicles via Fluorescent Nanoparticle Tracking Analysis

Yiting Liu[a], Anthony James El-helou[a], Bill Söderström[b], Juanfang Ruan[c] and Ying Zhu[*a,d,e,f]

[a] School of Biomedical Engineering, University of Technology Sydney, Australia

[b] Australian Institute for Microbiology and Infection, University of Technology Sydney, Australia

[c] Electron Microscope Unit, UNSW Sydney, Australia

[d] School of Clinical Medicine, Faculty of Medicine & Health, UNSW Sydney, Australia

[e] Institute for Biomedical Materials and Devices, University of Technology Sydney, Australia

[f] Australian Centre for NanoMedicine, UNSW Sydney, Australia

Table S1: Antibodies information used in this study. All antibodies' clonality is monoclonal.

| Target | Vendor | Cat No. | Clone | Host/Isotype |
|--------|--------|---------|-------|--------------|
| CD63 | Thermo Fisher Scientific | MA5-18149 | MEM-259 | Mouse/ IgG1 |
| CD63 | Abcam | ab309976 | EPR5702 | Rabbit/ IgG |
| ApoB | Santa Cruz | sc-393636 | A-6 | Mouse/ IgG1 |
| EpCAM | Abcam | ab237395 | EPR20532-225 | Rabbit/ IgG |

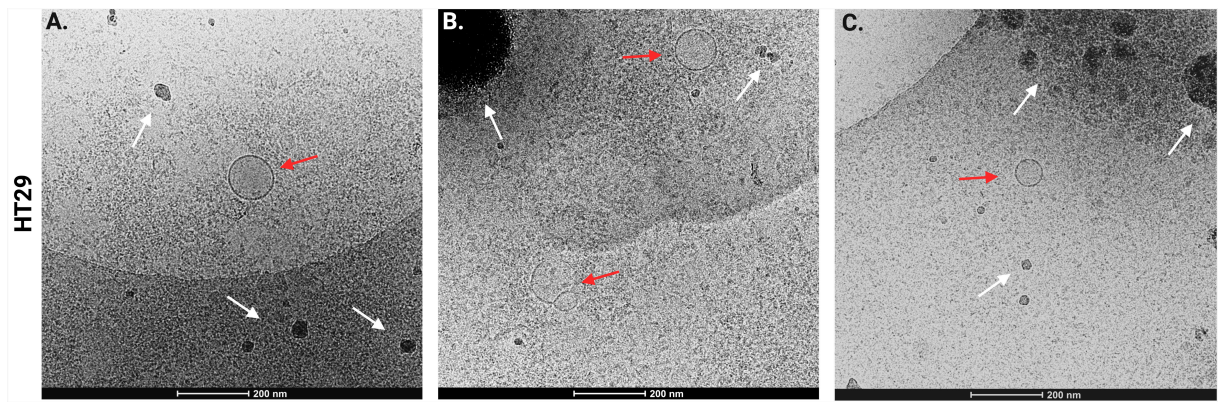

Figure S1. Representative Cryo-EM images of the commercially pursued HT29 EVs, showing the presence of EVs (red arrow) and other impurities (white arrow).

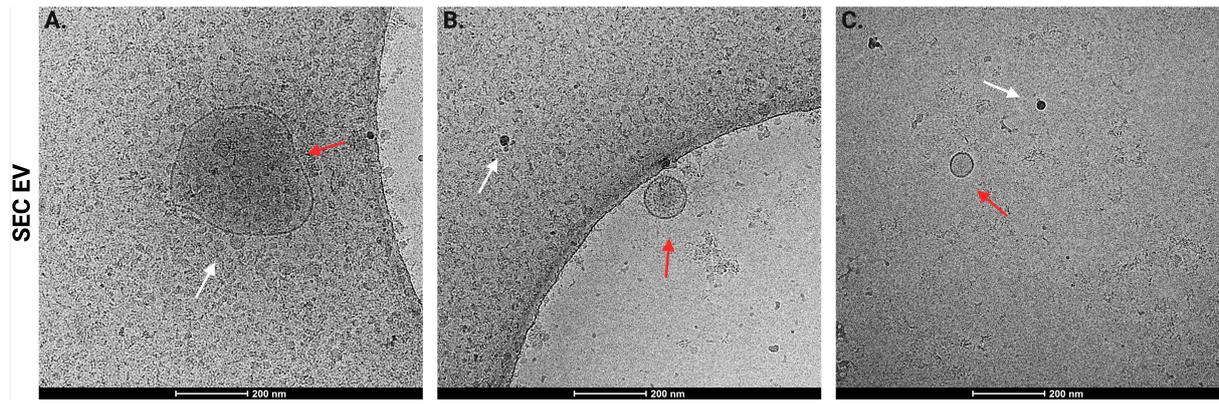

Figure S2. Representative Cryo-EM images of the EVs isolated from human plasma by SEC, showing the presence of EVs (red arrow) and other impurities (white arrow).

CD63 Freedye Fluo Concentration

| | F1 | F2 | F3 | F4 |
|---|---|---|---|---|
| P1 | 0.00E+00 | 0.00E+00 | 0.00E+00 | 0.00E+00 |
| P2 | 0.00E+00 | 0.00E+00 | 9.00E+05 | 9.00E+05 |
| P3 | 0.00E+00 | 0.00E+00 | 1.80E+06 | 9.00E+05 |
| P4 | 0.00E+00 | 9.00E+05 | 9.00E+05 | 0.00E+00 |
| P5 | 0.00E+00 | 0.00E+00 | 1.80E+06 | 0.00E+00 |
| P6 | 0.00E+00 | 1.80E+06 | 1.80E+06 | 1.80E+06 |
| P7 | 0.00E+00 | 1.80E+06 | 2.70E+06 | 0.00E+00 |
| P8 | 0.00E+00 | 0.00E+00 | 4.50E+06 | 9.00E+05 |
| P9 | 1.80E+06 | 2.70E+06 | 3.60E+06 | 1.80E+06 |
| P10 | 9.00E+05 | 9.00E+05 | 3.60E+06 | 1.80E+06 |
| P11 | 9.00E+05 | 1.80E+06 | 3.60E+06 | 0.00E+00 |

CD63 HT29 Fluo Conc

| | F1 |
|---|---|
| P1 | 8.10E+06 |
| P2 | 2.00E+07 |
| P3 | 2.50E+07 |
| P4 | 2.50E+07 |
| P5 | 3.50E+07 |
| P6 | 3.70E+07 |
| P7 | 2.10E+07 |
| P8 | 3.70E+07 |
| P9 | 2.00E+07 |
| P10 | 2.80E+07 |
| P11 | 2.90E+07 |

| | NVS F1 |
|---|---|
| P1 | 9.12E+06 |
| P2 | 2.25E+07 |
| P3 | 2.82E+07 |
| P4 | 2.82E+07 |
| P5 | 3.94E+07 |
| P6 | 4.17E+07 |
| P7 | 2.36E+07 |
| P8 | 4.17E+07 |
| P9 | 2.25E+07 |
| P10 | 3.15E+07 |
| P11 | 3.27E+07 |

CD63 HT29 Size

| | Scatter | | | | | 488nm | |
|---|---|---|---|---|---|---|---|
| | F1 | F2 | F3 | F4 | | F1 | F2 |
| P1 | 125 | 135 | 112.9 | 108.3 | P1 | 150 | 120 |
| P2 | 123.8 | 127.7 | 115.6 | 103.8 | P2 | 180 | 135 |
| P3 | 121.5 | 127.3 | 114.3 | 103.5 | P3 | 135 | 145 |
| P4 | 116.2 | 125.5 | 119.1 | 108.5 | P4 | 150 | 135 |
| P5 | 120.6 | 122.9 | 109.3 | 107.1 | P5 | 125 | 132.1 |
| P6 | 121.2 | 123.8 | 105 | 106 | P6 | 130 | 118.3 |
| P7 | 55 | 123.4 | 78.3 | 95 | P7 | 137.5 | 122.5 |
| P8 | 126.7 | 122 | 107.6 | 95.5 | P8 | 140 | 125 |
| P9 | 128.6 | 129 | 129 | 106.3 | P9 | 150 | 131 |
| P10 | 132.5 | 129.5 | 106.8 | 114.2 | P10 | 132.5 | 130 |
| P11 | 111.7 | 126.9 | 126.9 | 108 | P11 | 167.5 | 150 |

AopB Freedye Fluo Concentration

| | F1 | F2 | F3 | F4 |
|---|---|---|---|---|
| P1 | 4.50E+06 | 1.60E+07 | 1.10E+07 | 9.90E+06 |
| P2 | 7.20E+06 | 1.40E+07 | 1.20E+07 | 1.10E+07 |
| P3 | 8.10E+06 | 1.30E+07 | 7.20E+06 | 9.90E+06 |
| P4 | 9.00E+06 | 1.30E+07 | 9.00E+06 | 1.30E+07 |
| P5 | 8.10E+06 | 2.10E+07 | 7.20E+06 | 1.30E+07 |

Lipoprotein Fluo Conc

| | F1 | F2 |
|---|---|---|
| P1 | 9.00E+06 | 7.70E+07 |
| P2 | 9.00E+06 | 9.50E+07 |
| P3 | 1.20E+07 | 8.30E+07 |
| P4 | 8.10E+06 | 6.80E+07 |
| P5 | 4.50E+06 | 9.20E+07 |

| | | | | | | | |
|----|----------|----------|----------|----------|----|----------|----------|
| P6 | 7.20E+06 | 2.70E+07 | 1.10E+07 | 7.20E+06 | P6 | 1.30E+07 | 8.80E+07 |
| P7 | 8.10E+06 | 1.50E+07 | 9.90E+06 | 1.20E+07 | P7 | 1.70E+07 | 9.90E+07 |
| P8 | 9.90E+06 | 2.10E+07 | 1.10E+07 | 1.30E+07 | P8 | 5.40E+06 | 1.00E+08 |
| P9 | 9.00E+06 | 2.60E+07 | 1.90E+07 | 3.00E+07 | P9 | 1.10E+07 | 8.80E+07 |
| P10 | 9.00E+06 | 2.60E+07 | 1.80E+07 | 1.70E+07 | P10 | 1.40E+07 | 1.00E+08 |
| P11 | 8.10E+06 | 2.60E+07 | 1.30E+07 | 1.40E+07 | P11 | 9.90E+06 | 9.00E+07 |

liposomes Fluo Concentration

| | F1 | F2 | F3 | F4 |
|-----|----------|----------|----------|----------|
| P1 | 2.00E+07 | 4.40E+07 | 1.30E+07 | 3.60E+06 |
| P2 | 1.60E+07 | 4.40E+07 | 1.90E+07 | 4.50E+06 |
| P3 | 1.90E+07 | 4.80E+07 | 2.10E+07 | 7.20E+06 |
| P4 | 1.50E+07 | 3.90E+07 | 1.30E+07 | 7.20E+06 |
| P5 | 9.90E+06 | 4.00E+07 | 2.30E+07 | |
| P6 | 2.10E+07 | 5.10E+07 | 2.50E+07 | |
| P7 | 1.70E+07 | 4.10E+07 | 2.70E+07 | 5.40E+06 |
| P8 | 1.30E+07 | 2.80E+07 | 1.30E+07 | 3.60E+06 |
| P9 | 2.50E+07 | 5.20E+07 | 2.20E+07 | 5.40E+06 |
| P10 | 1.80E+07 | 4.10E+07 | 2.40E+07 | |
| P11 | 1.30E+07 | 3.60E+07 | 1.30E+07 | 6.30E+06 |

liposomes Scatter Concentration

| | F1 | F2 |
|-----|----------|----------|
| P1 | | 4.80E+07 |
| P2 | 3.00E+07 | 5.00E+07 |
| P3 | 2.50E+07 | 5.40E+07 |
| P4 | | 5.00E+07 |
| P5 | 2.90E+07 | 6.30E+07 |
| P6 | 3.00E+07 | 6.10E+07 |
| P7 | 2.40E+07 | 6.00E+07 |
| P8 | 3.00E+07 | 6.30E+07 |
| P9 | 3.00E+07 | 6.90E+07 |
| P10 | 3.70E+07 | 5.90E+07 |
| P11 | 5.40E+07 | 6.60E+07 |

| NVS F1 | NVS F2 | NVS F3 | NVS F4 |
|----------|----------|----------|----------|
| 2.04E+07 | 4.49E+07 | 1.33E+07 | 3.67E+06 |
| 1.63E+07 | 4.49E+07 | 1.94E+07 | 4.59E+06 |
| 1.94E+07 | 4.90E+07 | 2.14E+07 | 7.34E+06 |
| 1.53E+07 | 3.98E+07 | 1.33E+07 | 7.34E+06 |
| 1.01E+07 | 4.08E+07 | 2.35E+07 | |
| 2.14E+07 | 5.20E+07 | 2.55E+07 | |
| 1.73E+07 | 4.18E+07 | 2.75E+07 | 5.51E+06 |
| 1.33E+07 | 2.86E+07 | 1.33E+07 | 3.67E+06 |
| 2.55E+07 | 5.30E+07 | 2.24E+07 | 5.51E+06 |
| 1.84E+07 | 4.18E+07 | 2.45E+07 | |
| 1.33E+07 | 3.67E+07 | 1.33E+07 | 6.43E+06 |

Liposomes Strep Size

| | Scatter | | | | | 488nm | |
|-----|---------|-------|-------|-------|-----|-------|-------|
| | F1 | F2 | F3 | F4 | | F1 | F2 |
| P1 | 123.8 | 111.4 | 113.3 | 113.3 | P1 | 105 | 121 |
| P2 | 109.2 | 113.8 | 96.2 | 96.2 | P2 | 102.5 | 112 |
| P3 | 116.7 | 120 | 117.1 | 117.1 | P3 | 120 | 107 |
| P4 | 116.7 | 124.2 | 117.9 | 117.9 | P4 | 93.3 | 110 |
| P5 | 125 | 109.4 | 111.7 | 111.7 | P5 | 125 | 111.7 |
| P6 | 116.4 | 112.5 | 113 | 113 | P6 | 117.5 | 120 |
| P7 | 120 | 124.2 | 115 | 115 | P7 | 130 | 127.5 |
| P8 | 115 | 114 | 116.4 | 116.4 | P8 | 105 | 110 |
| P9 | 112 | 133 | 103.8 | 103.8 | P9 | 115 | 112 |
| P10 | 111 | 113.8 | 113.8 | 113.8 | P10 | 125 | 112.5 |
| P11 | 92.5 | 117.5 | 126.2 | 126.2 | P11 | 145 | 128.3 |

| F2 | F3 | F4 | MEAN | NVS | | CD63 HT29 Scatter Co |
|---|---|---|---|---|---|---|
| | | | | | | F1 |
| 8.50E+07 | 4.50E+07 | 1.30E+07 | 3.78E+07 | 4.25E+07 | P1 | 3.60E+08 |
| 9.00E+07 | 4.70E+07 | 2.10E+07 | 4.45E+07 | 5.01E+07 | P2 | 4.10E+08 |
| 1.00E+08 | 5.40E+07 | 2.50E+07 | 5.10E+07 | 5.74E+07 | P3 | 3.70E+08 |
| 1.20E+08 | 4.70E+07 | | 6.40E+07 | 7.21E+07 | P4 | 3.80E+08 |
| 1.30E+08 | 4.50E+07 | 3.20E+07 | 6.05E+07 | 6.81E+07 | P5 | 3.50E+08 |
| 1.20E+08 | 6.80E+07 | | 7.50E+07 | 8.45E+07 | P6 | 3.90E+08 |
| 6.80E+07 | 4.30E+07 | 1.30E+07 | 3.63E+07 | 4.08E+07 | P7 | |
| 1.20E+08 | 5.70E+07 | 1.80E+07 | 5.80E+07 | 6.53E+07 | P8 | 3.90E+08 |
| 9.60E+07 | 5.60E+07 | 2.50E+07 | 4.93E+07 | 5.55E+07 | P9 | 2.60E+08 |
| 3.20E+07 | 5.70E+07 | 3.20E+07 | 3.73E+07 | 4.19E+07 | P10 | 2.80E+07 |
| 6.60E+07 | 5.00E+07 | 2.50E+07 | 4.25E+07 | 4.79E+07 | P11 | 2.60E+08 |

| NVS F2 | NVS F3 | NVS F4 |
|---|---|---|
| 9.57E+07 | 5.07E+07 | 1.46E+07 |
| 1.01E+08 | 5.29E+07 | 2.36E+07 |
| 1.13E+08 | 6.08E+07 | 2.82E+07 |
| 1.35E+08 | 5.29E+07 | 0.00E+00 |
| 1.46E+08 | 5.07E+07 | 3.60E+07 |
| 1.35E+08 | 7.66E+07 | 0.00E+00 |
| 7.66E+07 | 4.84E+07 | 1.46E+07 |
| 1.35E+08 | 6.42E+07 | 2.03E+07 |
| 1.08E+08 | 6.31E+07 | 2.82E+07 |
| 3.60E+07 | 6.42E+07 | 3.60E+07 |
| 7.43E+07 | 5.63E+07 | 2.82E+07 |

| F3 | F4 |
|---|---|
| 112.5 | 140 |
| 170 | 85 |
| 140 | 226.7 |
| 112.5 | |
| 167.5 | 155 |
| 122.5 | |
| 107.5 | |
| 122.5 | 90 |
| 105 | 80 |
| 140 | 105 |
| 109 | 125 |

| F3 | F4 |
|---|---|
| 1.70E+08 | 1.90E+08 |
| 2.20E+08 | 2.10E+08 |
| 2.70E+08 | 2.50E+08 |
| 2.60E+08 | 3.20E+08 |
| 2.60E+08 | 2.90E+08 |
| 2.60E+08 | 3.40E+08 |
| 2.00E+08 | 2.50E+08 |
| 2.70E+08 | 3.20E+08 |
| 2.50E+08 | 3.00E+08 |
| 2.40E+08 | 2.60E+08 |
| 2.10E+08 | 2.60E+08 |

| F3 | F4 |
|---|---|
| 2.50E+07 | 1.20E+07 |
| 2.50E+07 | 1.30E+07 |
| 2.60E+07 | |
| 3.00E+07 | 9.90E+06 |
| 2.80E+07 | 1.40E+07 |
| 3.00E+07 | |
| 3.50E+07 | 1.40E+07 |
| 2.90E+07 | 1.50E+07 |
| 2.70E+07 | 1.50E+07 |
| 3.00E+07 | 1.50E+07 |
| 4.70E+07 | 3.90E+07 |

| F3 | F4 |
|---|---|
| 95 | 50 |
| 87.5 | 90 |
| 115 | 101.7 |
| 92.5 | 85 |
| 130 | 122.5 |
| 100 | 57.5 |
| 90 | 150 |
| 90 | 150 |
| 123 | 140 |
| 117.5 | 75 |
| 85 | 65 |

ncentration

| F2 | F3 | F4 | MEAN |
|---|---|---|---|
| 6.1E+08 | 6.9E+08 | 1.8E+08 | 4.60E+08 |
| 5.9E+08 | 7.4E+08 | 1.6E+08 | 4.75E+08 |
| 5E+08 | 6.2E+08 | 1.6E+08 | 4.13E+08 |
| 6E+08 | 6.6E+08 | 1.7E+08 | 4.53E+08 |
| 5.2E+08 | 6.7E+08 | 1.7E+08 | 4.28E+08 |
| 6E+08 | 6.9E+08 | 1.6E+08 | 4.60E+08 |
| 4.8E+08 | | | 4.80E+08 |
| 6.6E+08 | 7.1E+08 | 1.6E+08 | 4.80E+08 |
| 4.9E+08 | 9.8E+08 | 1.7E+08 | 4.75E+08 |
| 3.7E+08 | 6.9E+08 | 2.3E+08 | 3.30E+08 |
| 3.2E+08 | 6.3E+08 | 4.1E+08 | 4.05E+08 |

## HEK293 EPCAM Fluo

| | M1 | M2 | M3 |
|---|---|---|---|
| P1 | | 1.20E+07 | 1.10E+07 |
| P2 | 1.70E+07 | 1.50E+07 | 9.90E+06 |
| P3 | | 1.80E+07 | 1.20E+07 |
| P4 | 1.60E+07 | | |
| P5 | 1.30E+07 | 1.60E+07 | 1.10E+07 |
| P6 | 9.90E+06 | | 1.10E+07 |
| P7 | 1.30E+07 | 1.30E+07 | 1.30E+07 |
| P8 | | 1.40E+07 | 1.30E+07 |
| P9 | 1.60E+07 | 2.10E+07 | |
| P10 | 1.50E+07 | | 1.10E+07 |
| P11 | 1.40E+07 | 1.20E+07 | 6.30E+06 |

| | NVS M1 | NVS M2 | NVS M3 |
|---|---|---|---|
| P1 | | 2.40E+07 | 2.20E+07 |
| P2 | 3.40E+07 | 3.00E+07 | 1.98E+07 |
| P3 | | 3.60E+07 | 2.40E+07 |
| P4 | 3.20E+07 | | |
| P5 | 2.60E+07 | 3.20E+07 | 2.20E+07 |
| P6 | 1.98E+07 | | 2.20E+07 |
| P7 | 2.60E+07 | 2.60E+07 | 2.60E+07 |
| P8 | | 2.80E+07 | 2.60E+07 |
| P9 | 3.20E+07 | 4.20E+07 | |
| P10 | 3.00E+07 | | 2.20E+07 |
| P11 | 2.80E+07 | 2.40E+07 | 1.26E+07 |

## HEK293 EPCAM Scatter

| | M1 | M2 | M3 |
|---|---|---|---|
| P1 | 1.30E+08 | 1.30E+08 | 1.30E+08 |
| P2 | 1.30E+08 | 1.30E+08 | 1.50E+08 |
| P3 | 1.10E+08 | 1.10E+08 | 1.30E+08 |
| P4 | 1.10E+08 | 1.40E+08 | 1.30E+08 |
| P5 | 1.10E+08 | 1.10E+08 | 1.30E+08 |
| P6 | 1.10E+08 | 1.20E+08 | 1.30E+08 |
| P7 | 1.30E+08 | 1.20E+08 | 1.10E+08 |
| P8 | 1.40E+08 | | 1.60E+08 |
| P9 | | 1.70E+08 | 1.50E+08 |
| P10 | | | |
| P11 | 3.40E+08 | 3.60E+08 | 3.60E+08 |

## HT29 EPCAM Fluo

| | M1 | M2 | M3 |
|---|---|---|---|
| P1 | 1.30E+08 | 8.70E+07 | 9.10E+07 |
| P2 | 1.20E+08 | 1.10E+08 | 1.20E+08 |
| P3 | 9.90E+07 | 1.10E+08 | 1.20E+08 |
| P4 | 1.20E+08 | 9.90E+07 | 1.20E+08 |
| P5 | 9.00E+07 | 1.20E+08 | 1.30E+08 |
| P6 | 4.20E+07 | 1.10E+08 | 1.20E+08 |
| P7 | 8.90E+07 | 1.00E+08 | 1.10E+08 |
| P8 | 1.10E+08 | 1.10E+08 | 1.50E+08 |
| P9 | 1.00E+08 | 1.10E+08 | 1.40E+08 |
| P10 | 1.10E+08 | 9.50E+07 | 1.20E+08 |
| P11 | 6.70E+07 | 7.50E+07 | 1.00E+08 |

| | NVS M1 | NVS M2 | NVS M3 |
|---|---|---|---|
| P1 | 2.24E+08 | 1.50E+08 | 1.57E+08 |
| P2 | 2.07E+08 | 1.90E+08 | 2.07E+08 |
| P3 | 1.71E+08 | 1.90E+08 | 2.07E+08 |
| P4 | 2.07E+08 | 1.71E+08 | 2.07E+08 |
| P5 | 1.55E+08 | 2.07E+08 | 2.24E+08 |
| P6 | 7.25E+07 | 1.90E+08 | 2.07E+08 |
| P7 | 1.54E+08 | 1.73E+08 | 1.90E+08 |
| P8 | 1.90E+08 | 1.90E+08 | 2.59E+08 |
| P9 | 1.73E+08 | 1.90E+08 | 2.42E+08 |
| P10 | 1.90E+08 | 1.64E+08 | 2.07E+08 |

## HT29 EPCAM Scatter

| | M1 | M2 | M3 |
|---|---|---|---|
| P1 | 3.40E+08 | 3.30E+08 | 3.20E+08 |
| P2 | 3.30E+08 | 3.10E+08 | 3.20E+08 |
| P3 | 2.30E+08 | 2.50E+08 | 2.40E+08 |
| P4 | 2.50E+08 | 2.90E+08 | 2.70E+08 |
| P5 | 2.90E+08 | 2.70E+08 | 2.30E+08 |
| P6 | 2.90E+08 | 2.90E+08 | 2.60E+08 |
| P7 | 3.10E+08 | 3.10E+08 | 2.50E+08 |
| P8 | 2.70E+08 | 2.80E+08 | 2.90E+08 |
| P9 | 3.30E+08 | 3.60E+08 | 3.20E+08 |
| P10 | 3.90E+08 | 3.90E+08 | 4.30E+08 |
| P11 | 4.00E+08 | 4.00E+08 | 4.10E+08 |

P11     1.16E+08  1.29E+08  1.73E+08

**HEK293 Fluo to Scatter**

| | M1 | M2 | M3 |
|---|---|---|---|
| P1 | NA | 18% | 17% |
| P2 | 26% | 23% | 13% |
| P3 | NA | 33% | 18% |
| P4 | 29% | NA | NA |
| P5 | 24% | 29% | 17% |
| P6 | 18% | NA | 17% |
| P7 | 20% | 22% | 24% |
| P8 | NA | NA | 16% |
| P9 | NA | 25% | NA |
| P10 | NA | NA | NA |
| P11 | 8% | 7% | 4% |

**HT29 Fluo to Scatter**

| | M1 | M2 | M3 |
|---|---|---|---|
| P1 | 66% | 45% | 49% |
| P2 | 63% | 61% | 65% |
| P3 | 74% | 76% | 86% |
| P4 | 83% | 59% | 77% |
| P5 | 54% | 77% | 98% |
| P6 | 25% | 65% | 80% |
| P7 | 50% | 56% | 76% |
| P8 | 70% | 68% | 89% |
| P9 | 52% | 53% | 75% |
| P10 | 49% | 42% | 48% |
| P11 | 29% | 32% | 42% |

**HT29 Size**

| Size / nm | | Scatter |
|---|---|---|
| 5 | 0 | 70120 |
| 15 | 148900 | 210300 |
| 25 | 223400 | 490800 |
| 35 | 744700 | 771300 |
| 45 | 1936000 | 1262000 |
| 55 | 1713000 | 2384000 |
| 65 | 4617000 | 5469000 |
| 75 | 7447000 | 8274000 |
| 85 | 11620000 | 12270000 |
| 95 | 15120000 | 15570000 |
| 105 | 17650000 | 18510000 |
| 115 | 15490000 | 15290000 |
| 125 | 13930000 | 15220000 |
| 135 | 11620000 | 12130000 |
| 145 | 11990000 | 9606000 |
| 155 | 8564000 | 8554000 |
| 165 | 7522000 | 7993000 |
| 175 | 5511000 | 4628000 |
| 185 | 2904000 | 4417000 |
| 195 | 2607000 | 4137000 |
| 205 | 2160000 | 2454000 |
| 215 | 1564000 | 1543000 |
| 225 | 1787000 | 1402000 |
| 235 | 1564000 | 1332000 |
| 245 | 670300 | 701200 |
| 255 | 595800 | 490800 |
| 265 | 819200 | 841400 |
| 275 | 148900 | 420700 |
| 285 | 372400 | 280500 |
| 295 | 223400 | 70120 |
| 305 | 148900 | 140200 |
| 315 | 74470 | 210300 |
| 325 | 297900 | 70120 |
| 335 | 74470 | 70120 |
| 345 | 74470 | 0 |
| 355 | 148900 | 210300 |
| 365 | 74470 | 70120 |
| 375 | 74470 | 140200 |
| 385 | 0 | 0 |
| 395 | 0 | 70120 |
| 405 | 0 | 70120 |
| 415 | 74470 | 0 |
| 425 | 0 | 140200 |
| 435 | 0 | 0 |
| 445 | 0 | 70120 |
| 455 | 0 | 0 |
| 465 | 0 | 0 |
| 475 | 0 | 0 |
| 485 | 0 | 0 |
| 495 | 0 | 0 |
| 505 | 0 | 0 |
| 515 | 74470 | 0 |
| 525 | 0 | 0 |
| 535 | 0 | 0 |
| 545 | 0 | 0 |
| 555 | 74470 | 0 |
| 565 | 0 | 0 |
| 575 | 0 | 0 |
| 585 | 0 | 0 |
| 595 | 0 | 0 |

HEK293 Size

| 488nm | | | | Size / nm | Scatter | | |
|---|---|---|---|---|---|---|---|
| 0 | 0 | 0 | 80280 | 0 | | | |
| 591700 | 0 | 0 | 0 | 5 | 0 | 0 | 0 |
| 591700 | 142100 | 73320 | 80280 | 15 | 69200 | 72660 | 197100 |
| 739600 | 355200 | 146600 | 642200 | 25 | 346000 | 72660 | 0 |
| 1553000 | 639400 | 586600 | 963400 | 35 | 138400 | 363300 | 131400 |
| 2663000 | 994700 | 2200000 | 2007000 | 45 | 0 | 290600 | 262700 |
| 5029000 | 1563000 | 2786000 | 2007000 | 55 | 761200 | 726600 | 394100 |
| 8727000 | 3197000 | 2493000 | 4335000 | 65 | 830400 | 1090000 | 1051000 |
| 12720000 | 2771000 | 3373000 | 4415000 | 75 | 2007000 | 2834000 | 2365000 |
| 14720000 | 4405000 | 4106000 | 4977000 | 85 | 3391000 | 4287000 | 5452000 |
| 15010000 | 4334000 | 4106000 | 4897000 | 95 | 4636000 | 5522000 | 6634000 |
| 14640000 | 3908000 | 5792000 | 4014000 | 105 | 6850000 | 6830000 | 7751000 |
| 13900000 | 4192000 | 3593000 | 5459000 | 115 | 9549000 | 8210000 | 8013000 |
| 13170000 | 3695000 | 3446000 | 4255000 | 125 | 8165000 | 8646000 | 8473000 |
| 10130000 | 2416000 | 2860000 | 3773000 | 135 | 6850000 | 8138000 | 6897000 |
| 8210000 | 3055000 | 2346000 | 2248000 | 145 | 5190000 | 6176000 | 5846000 |
| 6361000 | 2558000 | 2566000 | 2168000 | 155 | 5397000 | 5086000 | 5058000 |
| 5547000 | 1776000 | 1320000 | 1846000 | 165 | 4083000 | 3488000 | 3744000 |
| 4068000 | 1847000 | 1320000 | 1766000 | 175 | 2768000 | 2034000 | 2890000 |
| 4216000 | 994700 | 733200 | 1606000 | 185 | 2560000 | 2761000 | 2562000 |
| 2071000 | 781500 | 439900 | 1044000 | 195 | 1453000 | 2034000 | 1708000 |
| 1331000 | 923600 | 1320000 | 883100 | 205 | 1176000 | 1889000 | 1576000 |
| 1183000 | 994700 | 879900 | 802800 | 215 | 1176000 | 1453000 | 1642000 |
| 1183000 | 568400 | 806500 | 802800 | 225 | 1176000 | 944600 | 1182000 |
| 739600 | 355200 | 659900 | 562000 | 235 | 553600 | 944600 | 722500 |
| 295800 | 284200 | 513300 | 562000 | 245 | 761200 | 363300 | 394100 |
| 517700 | 426300 | 366600 | 802800 | 255 | 553600 | 508600 | 591200 |
| 369800 | 213100 | 220000 | 321100 | 265 | 276800 | 726600 | 525500 |
| 443800 | 568400 | 220000 | 401400 | 275 | 138400 | 72660 | 459800 |
| 147900 | 213100 | 146600 | 642200 | 285 | 553600 | 290600 | 394100 |
| 147900 | 142100 | 366600 | 160600 | 295 | 276800 | 145300 | 328400 |
| 0 | 71050 | 73320 | 321100 | 305 | 276800 | 363300 | 65680 |
| 147900 | 213100 | 73320 | 160600 | 315 | 69200 | 145300 | 131400 |
| 0 | 0 | 146600 | 240800 | 325 | 138400 | 363300 | 197100 |
| 0 | 142100 | 220000 | 0 | 335 | 207600 | 72660 | 65680 |
| 0 | 0 | 146600 | 80280 | 345 | 69200 | 72660 | 131400 |
| 0 | 71050 | 146600 | 160600 | 355 | 69200 | 72660 | 0 |
| 0 | 0 | 0 | 0 | 365 | 138400 | 72660 | 65680 |
| 73960 | 0 | 0 | 0 | 375 | 69200 | 0 | 0 |
| 73960 | 142100 | 73320 | 80280 | 385 | 0 | 0 | 0 |
| 73960 | 142100 | 0 | 240800 | 395 | 0 | 72660 | 0 |
| 0 | 71050 | 0 | 0 | 405 | 138400 | 72660 | 0 |
| 0 | 71050 | 73320 | 0 | 415 | 0 | 72660 | 131400 |
| 0 | 0 | 0 | 80280 | 425 | 69200 | 0 | 0 |
| 0 | 71050 | 73320 | 0 | 435 | 0 | 0 | 0 |
| 0 | 0 | 0 | 0 | 445 | 0 | 72660 | 0 |
| 73960 | 0 | 0 | 80280 | 455 | 0 | 0 | 0 |
| 0 | 0 | 0 | 0 | 465 | 0 | 145300 | 0 |
| 0 | 0 | 0 | 0 | 475 | 69200 | 0 | 0 |
| 0 | 0 | 0 | 0 | 485 | 0 | 0 | 0 |
| 0 | 0 | 0 | 0 | 495 | 69200 | 0 | 0 |
| 0 | 0 | 0 | 0 | 505 | 0 | 0 | 0 |
| 0 | 0 | 0 | 0 | 515 | 0 | 0 | 65680 |
| 0 | 0 | 0 | 0 | 525 | 0 | 0 | 0 |
| 0 | 0 | 0 | 0 | 535 | 0 | 0 | 0 |
| 0 | 0 | 0 | 0 | 545 | 0 | 0 | 0 |
| 0 | 0 | 0 | 0 | 555 | 0 | 0 | 0 |
| 0 | 0 | 0 | 0 | 565 | 0 | 0 | 0 |
| 0 | 0 | 0 | 0 | 575 | 0 | 0 | 0 |
| 0 | 0 | 0 | 0 | 585 | 0 | 0 | 0 |
| | | | | 595 | 0 | 0 | 0 |

488nm

| | | |
|---|---|---|
| 0 | 0 | 0 |
| 0 | 0 | 0 |
| 0 | 0 | 0 |
| 242900 | 225600 | 479800 |
| 971500 | 676700 | 575700 |
| 485700 | 676700 | 479800 |
| 850100 | 676700 | 575700 |
| 485700 | 338400 | 863600 |
| 728600 | 563900 | 671700 |
| 607200 | 451100 | 287900 |
| 485700 | 563900 | 191900 |
| 485700 | 338400 | 95960 |
| 121400 | 563900 | 191900 |
| 364300 | 338400 | 0 |
| 0 | 451100 | 191900 |
| 364300 | 563900 | 191900 |
| 0 | 0 | 0 |
| 121400 | 112800 | 0 |
| 0 | 338400 | 0 |
| 121400 | 112800 | 95960 |
| 121400 | 0 | 95960 |
| 121400 | 112800 | 0 |
| 0 | 0 | 0 |
| 121400 | 112800 | 95960 |
| 0 | 0 | 0 |
| 0 | 0 | 95960 |
| 0 | 0 | 0 |
| 121400 | 0 | 95960 |
| 121400 | 0 | 0 |
| 0 | 0 | 95960 |
| 0 | 0 | 0 |
| 0 | 0 | 0 |
| 0 | 0 | 0 |
| 0 | 0 | 0 |
| 0 | 112800 | 0 |
| 121400 | 0 | 0 |
| 0 | 0 | 0 |
| 0 | 0 | 0 |
| 0 | 0 | 0 |
| 0 | 0 | 0 |
| 0 | 0 | 0 |
| 0 | 0 | 0 |
| 0 | 0 | 0 |
| 0 | 0 | 0 |
| 0 | 0 | 0 |
| 0 | 0 | 0 |
| 0 | 0 | 0 |
| 0 | 0 | 0 |
| 0 | 0 | 0 |
| 0 | 0 | 0 |
| 0 | 0 | 0 |
| 0 | 112800 | 0 |
| 0 | 0 | 0 |
| 0 | 0 | 0 |
| 0 | 0 | 0 |
| 0 | 0 | 0 |
| 0 | 0 | 0 |
| 0 | 0 | 0 |
| 0 | 0 | 0 |
| 0 | 0 | 0 |

SEC EVs

## 488 nm

| | M1 | M2 | M3 |
|---|---|---|---|
| P1 | | 3.20E+08 | |
| P2 | 5.70E+08 | 6.00E+08 | 5.30E+08 |
| P3 | 7.00E+08 | 6.60E+08 | 5.80E+08 |
| P4 | 8.00E+08 | 7.60E+08 | 7.60E+08 |
| P5 | 7.50E+08 | 7.70E+08 | 7.00E+08 |
| P6 | 3.30E+08 | 7.60E+08 | 7.70E+08 |
| P7 | 8.10E+08 | 8.60E+08 | 8.10E+08 |
| P8 | 7.10E+08 | 7.60E+08 | 9.30E+08 |
| P9 | 7.20E+08 | 7.90E+08 | 8.80E+08 |
| P10 | 8.00E+08 | 7.10E+08 | 7.60E+08 |
| P11 | 6.90E+08 | 6.70E+08 | 7.50E+08 |

## 647 nm

| | M1 | M2 |
|---|---|---|
| P1 | 2.70E+07 | 1.80E+07 |
| P2 | 3.60E+07 | 3.60E+07 |
| P3 | 3.60E+07 | 4.50E+07 |
| P4 | 5.80E+07 | 4.90E+07 |
| P5 | 6.30E+07 | 3.60E+07 |
| P6 | 9.90E+07 | 5.40E+07 |
| P7 | 7.20E+07 | 8.50E+07 |
| P8 | 6.70E+07 | 7.20E+07 |
| P9 | 6.70E+07 | 7.20E+07 |
| P10 | 8.10E+07 | 7.20E+07 |
| P11 | 8.50E+07 | 9.40E+07 |

| 488 | S | F | NVS |
|---|---|---|---|
| | 142.00 | 125.00 | 1.136 |

| 647 | S | F |
|---|---|---|
| | 109.00 | 117.00 |

| | M1 NVS | M2 NVS | M3 NVS |
|---|---|---|---|
| P1 | | 3.64E+08 | |
| P2 | 6.48E+08 | 6.82E+08 | 6.02E+08 |
| P3 | 7.95E+08 | 7.50E+08 | 6.59E+08 |
| P4 | 9.09E+08 | 8.63E+08 | 8.63E+08 |
| P5 | 8.52E+08 | 8.75E+08 | 7.95E+08 |
| P6 | 3.75E+08 | 8.63E+08 | 8.75E+08 |
| P7 | 9.20E+08 | 9.77E+08 | 9.20E+08 |
| P8 | 8.07E+08 | 8.63E+08 | 1.06E+09 |
| P9 | 8.18E+08 | 8.97E+08 | 1.00E+09 |
| P10 | 9.09E+08 | 8.07E+08 | 8.63E+08 |
| P11 | 7.84E+08 | 7.61E+08 | 8.52E+08 |

| | M1 NVS | M2 NVS |
|---|---|---|
| P1 | 2.52E+07 | 1.68E+07 |
| P2 | 3.36E+07 | 3.36E+07 |
| P3 | 3.36E+07 | 4.19E+07 |
| P4 | 5.41E+07 | 4.57E+07 |
| P5 | 5.87E+07 | 3.36E+07 |
| P6 | 9.23E+07 | 5.03E+07 |
| P7 | 6.71E+07 | 7.92E+07 |
| P8 | 6.24E+07 | 6.71E+07 |
| P9 | 6.24E+07 | 6.71E+07 |
| P10 | 7.55E+07 | 6.71E+07 |
| P11 | 7.92E+07 | 8.76E+07 |

| | | | | | Size / nm | |
|---|---|---|---|---|---|---|
| | | | | | 5 | 0 |
| M3 | | | | | 15 | 147907.2 |
| | | | | | 25 | 1626752 |
| 3.10E+07 | | | | | 35 | 3401184 |
| 4.50E+07 | | | | | 45 | 6211648 |
| 6.70E+07 | | | | | 55 | 7690720 |
| 5.80E+07 | | | | | 65 | 6951184 |
| 8.50E+07 | | | | | 75 | 5472112 |
| 6.30E+07 | | | | | 85 | 4140720 |
| 9.00E+07 | | | | | 95 | 3697680 |
| 4.50E+07 | | | | | 105 | 3105824 |
| 7.20E+07 | | | | | 115 | 1922112 |
| 8.10E+07 | | | | | 125 | 1035237 |
| | | | | | 135 | 1922112 |
| NVS | | | | | 145 | 2070928 |
| 0.931624 | | | | | 155 | 1479072 |
| | | EV to NVEPs | | | 165 | 887329.6 |
| M3 NVS | | M1 | M2 | M3 | 175 | 2070928 |
| 0 | | | 4.61% | | 185 | 1331392 |
| 2.89E+07 | | 5.18% | 4.92% | 4.80% | 195 | 2070928 |
| 4.19E+07 | | 4.22% | 5.59% | 6.37% | 205 | 887329.6 |
| 6.24E+07 | | 5.95% | 5.29% | 7.23% | 215 | 1183712 |
| 5.41E+07 | | 6.89% | 3.84% | 6.80% | 225 | 739422.4 |
| 7.92E+07 | | 24.61% | 5.83% | 9.06% | 235 | 1183712 |
| 5.87E+07 | | 7.29% | 8.11% | 6.38% | 245 | 1479072 |
| 8.39E+07 | | 7.74% | 7.77% | 7.94% | 255 | 887329.6 |
| 4.19E+07 | | 7.63% | 7.48% | 4.20% | 265 | 1331392 |
| 6.71E+07 | | 8.31% | 8.32% | 7.77% | 275 | 443721.6 |
| 7.55E+07 | | 10.11% | 11.51% | 8.86% | 285 | 591628.8 |
| | | | | | 295 | 739422.4 |
| | | | | | 305 | 1183712 |
| | | | | | 315 | 295814.4 |
| | | | | | 325 | 739422.4 |
| | | | | | 335 | 0 |
| | | | | | 345 | 739422.4 |
| | | | | | 355 | 443721.6 |
| | | | | | 365 | 443721.6 |
| | | | | | 375 | 739422.4 |
| | | | | | 385 | 739422.4 |
| | | | | | 395 | 591628.8 |
| | | | | | 405 | 295814.4 |
| | | | | | 415 | 443721.6 |
| | | | | | 425 | 147907.2 |
| | | | | | 435 | 591628.8 |
| | | | | | 445 | 295814.4 |
| | | | | | 455 | 147907.2 |
| | | | | | 465 | 443721.6 |
| | | | | | 475 | 147907.2 |
| | | | | | 485 | 0 |

| | |
|---|---|
| 495 | 295814.4 |
| 505 | 0 |
| 515 | 147907.2 |
| 525 | 0 |
| 535 | 147907.2 |
| 545 | 0 |
| 555 | 147907.2 |
| 565 | 0 |
| 575 | 295814.4 |
| 585 | 0 |
| 595 | 147907.2 |

| NVEPs | | EVs | | |
|---|---|---|---|---|
| 0 | 0 | 0 | 0 | 0 |
| 154609.6 | 153814.4 | 0 | 0 | 0 |
| 1700592 | 1691504 | 0 | 0 | 0 |
| 3555680 | 3537504 | 535217.9 | 580494.9 | 522827.3 |
| 6492240 | 6460432 | 535217.9 | 580494.9 | 522827.3 |
| 8037200 | 7998576 | 401436.7 | 435347.8 | 392120.5 |
| 7264720 | 7229504 | 669092.3 | 725641.9 | 653534.2 |
| 5718624 | 5691360 | 401436.7 | 435347.8 | 392120.5 |
| 4328160 | 4306576 | 267655.5 | 290294 | 261413.7 |
| 3864672 | 3845360 | 133781.2 | 145147 | 130706.8 |
| 3245552 | 3229648 | 535217.9 | 580494.9 | 522827.3 |
| 2009584 | 1999360 | 267655.5 | 290294 | 261413.7 |
| 1082040 | 1076701 | 401436.7 | 435347.8 | 392120.5 |
| 2009584 | 1999360 | 267655.5 | 290294 | 261413.7 |
| 2164080 | 2153856 | 0 | 0 | 0 |
| 1546096 | 1538144 | 133781.2 | 145147 | 130706.8 |
| 927430.4 | 922886.4 | 133781.2 | 145147 | 130706.8 |
| 2164080 | 2153856 | 133781.2 | 145147 | 130706.8 |
| 1391600 | 1384784 | 0 | 0 | 0 |
| 2164080 | 2153856 | 133781.2 | 145147 | 130706.8 |
| 927430.4 | 922886.4 | 133781.2 | 145147 | 130706.8 |
| 1237104 | 1230288 | 0 | 0 | 0 |
| 772820.8 | 769072 | 133781.2 | 145147 | 130706.8 |
| 1237104 | 1230288 | 0 | 0 | 0 |
| 1546096 | 1538144 | 267655.5 | 290294 | 261413.7 |
| 927430.4 | 922886.4 | 133781.2 | 145147 | 130706.8 |
| 1391600 | 1384784 | 0 | 0 | 0 |
| 463715.2 | 461443.2 | 133781.2 | 145147 | 130706.8 |
| 618324.8 | 615257.6 | 133781.2 | 145147 | 130706.8 |
| 772820.8 | 769072 | 0 | 0 | 0 |
| 1237104 | 1230288 | 267655.5 | 290294 | 261413.7 |
| 309105.6 | 307628.8 | 133781.2 | 145147 | 130706.8 |
| 772820.8 | 769072 | 267655.5 | 290294 | 261413.7 |
| 0 | 0 | 0 | 0 | 0 |
| 772820.8 | 769072 | 0 | 0 | 0 |
| 463715.2 | 461443.2 | 133781.2 | 145147 | 130706.8 |
| 463715.2 | 461443.2 | 0 | 0 | 0 |
| 772820.8 | 769072 | 267655.5 | 290294 | 261413.7 |
| 772820.8 | 769072 | 0 | 0 | 0 |
| 618324.8 | 615257.6 | 0 | 0 | 0 |
| 309105.6 | 307628.8 | 133781.2 | 145147 | 130706.8 |
| 463715.2 | 461443.2 | 0 | 0 | 0 |
| 154609.6 | 153814.4 | 0 | 0 | 0 |
| 618324.8 | 615257.6 | 0 | 0 | 0 |
| 309105.6 | 307628.8 | 133781.2 | 145147 | 130706.8 |
| 154609.6 | 153814.4 | 0 | 0 | 0 |
| 463715.2 | 461443.2 | 0 | 0 | 0 |
| 154609.6 | 153814.4 | 133781.2 | 145147 | 130706.8 |
| 0 | 0 | 0 | 0 | 0 |
| 309105.6 | 307628.8 | 0 | 0 | 0 |
| 0 | 0 | 0 | 0 | 0 |
| 154609.6 | 153814.4 | 0 | 0 | 0 |
| 0 | 0 | 0 | 0 | 0 |
| 154609.6 | 153814.4 | 0 | 0 | 0 |
| 0 | 0 | 0 | 0 | 0 |
| 154609.6 | 153814.4 | 0 | 0 | 0 |
| 0 | 0 | 0 | 0 | 0 |
| 309105.6 | 307628.8 | 0 | 0 | 0 |
| 0 | 0 | 0 | 0 | 0 |
| 154609.6 | 153814.4 | 0 | 0 | 0 |